\input{aipcheck}

\documentclass[
    ,final            
  ]
  {aipproc}

\layoutstyle{6x9}
\usepackage{natbib}

\begin{document}

\title{The unusual X-ray morphology of NGC~4636 revealed by deep Chandra observations: cavities and shocks created
by past AGN outbursts}


\author{A. Baldi}{
  address={Harvard-Smithsonian Center for Astrophysics, 60 Garden St, Cambridge, MA
  02138}
  ,altaddress={now at INAF-Osservatorio Astronomico di Bologna, Via Ranzani 1, 
  I-40127 Bologna, Italy} 
}

\author{W. Forman}{
  address={Harvard-Smithsonian Center for Astrophysics, 60 Garden St, Cambridge, MA 02138}
}

\author{C. Jones}{
  address={Harvard-Smithsonian Center for Astrophysics, 60 Garden St, Cambridge, MA 02138}
}

\author{R. Kraft}{
  address={Harvard-Smithsonian Center for Astrophysics, 60 Garden St, Cambridge, MA 02138}
}

\author{P. Nulsen}{
  address={Harvard-Smithsonian Center for Astrophysics, 60 Garden St, Cambridge, MA 02138}
}

\author{L. David}{
  address={Harvard-Smithsonian Center for Astrophysics, 60 Garden St, Cambridge, MA 02138}
}

\author{S. Giacintucci}{
  address={Harvard-Smithsonian Center for Astrophysics, 60 Garden St, Cambridge, MA 02138}
}

\author{E. Churazov}{
  address={Max Planck Institut f\"ur Astrophysik, Karl-Schwarzschild-Str. 1, 85741
  Garching, Germany}
}

\begin{abstract}
We present Chandra ACIS-I and ACIS-S observations ($\sim$200~ks in total) of the X-ray luminous 
elliptical 
galaxy NGC 4636, located in the outskirts of the Virgo cluster.
A soft band (0.5-2 keV) image shows the presence of a bright core in the center surrounded by an 
extended 
X-ray corona and two pronounced quasi-symmetric, 8 kpc long, arm-like features. Each of this 
features defines the rim
of an ellipsoidal bubble. An additional bubble-like feature, whose northern rim is located 
$\sim2$~kpc south of the 
north-eastern arm, is detected as well.
We present surface brightness and temperature profiles across the rims of the bubbles, showing 
that their edges are sharp 
and characterized by temperature jumps of about 20-25\%. 
Through a comparison of the observed profiles with theoretical shock models, we demonstrate that 
a scenario 
where the bubbles were produced by shocks, probably driven by energy deposited off-center by jets, 
is the most 
viable explanation to the X-ray morphology observed in the central part of NGC 4636.
\end{abstract}

\maketitle


\section{Introduction}

It is well known that active galactic nuclei (AGN) play an important role in the
evolution of the hot gas in both individual galaxies and clusters of galaxies
\citep[e.g. ][]{churazov00,bruggen02,cavaliere02,hoeft04,forman05,mcnamara07}.
The so-called `AGN feedback' can cause re-heating of the central cooling regions
of a cluster and balance the cooling due to the X-ray emission \citep[see][for a review]{mcnamara07}.
AGN are also effective in shaping the morphology of the hot gas halos around individual
galaxies and clusters, e.g. giving rise to cavities and bubbles \citep{birzan04}.\\
While several examples of subsonic bubble inflation due to AGN outbursts in individual galaxies are present
in the literature \citep[e.g. NGC~507, M~84; ][]{kraft04,finoguenov08}, supersonic bubble expansion,
giving rise to shocks in the hot X-ray emitting gas, has only been observed in a handful of cases
\citep[e.g. M~87, NGC~4552; ][]{forman05,machachek06}.
Another possible case of supersonic bubble expansion could be that of NGC 4636, 
the dominant galaxy of a group on the outskirts of the Virgo Cluster 
\citep[10$^\circ$ or 2.6 Mpc on the sky to the south of M87, at a distance to NGC~4636 of 15~Mpc;][]{tonry01}.

\section{Chandra Data Preparation and X-ray Morphology} \label{dataprep}

\begin{figure}
\includegraphics[width=6cm]{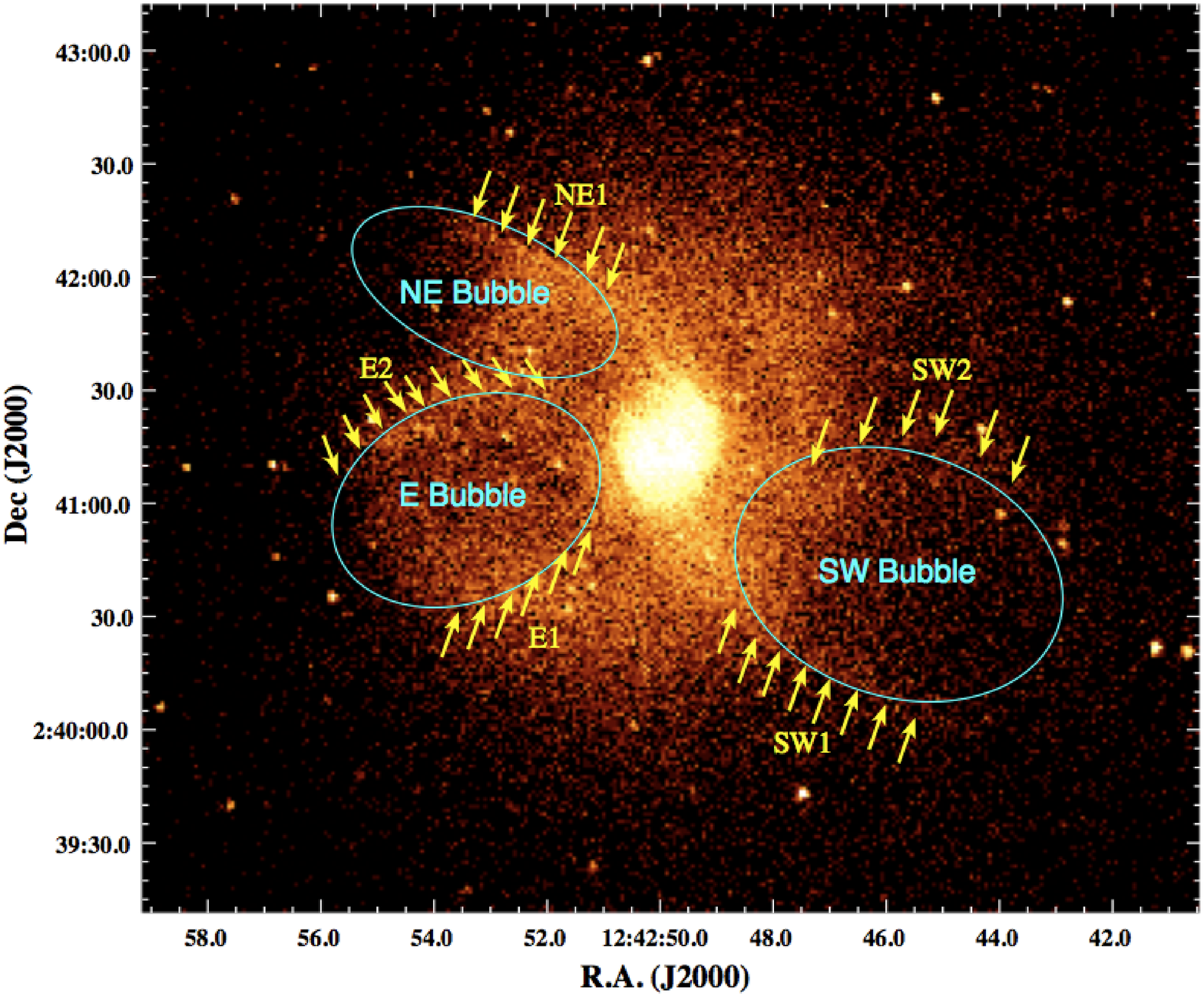}
\includegraphics[width=6cm]{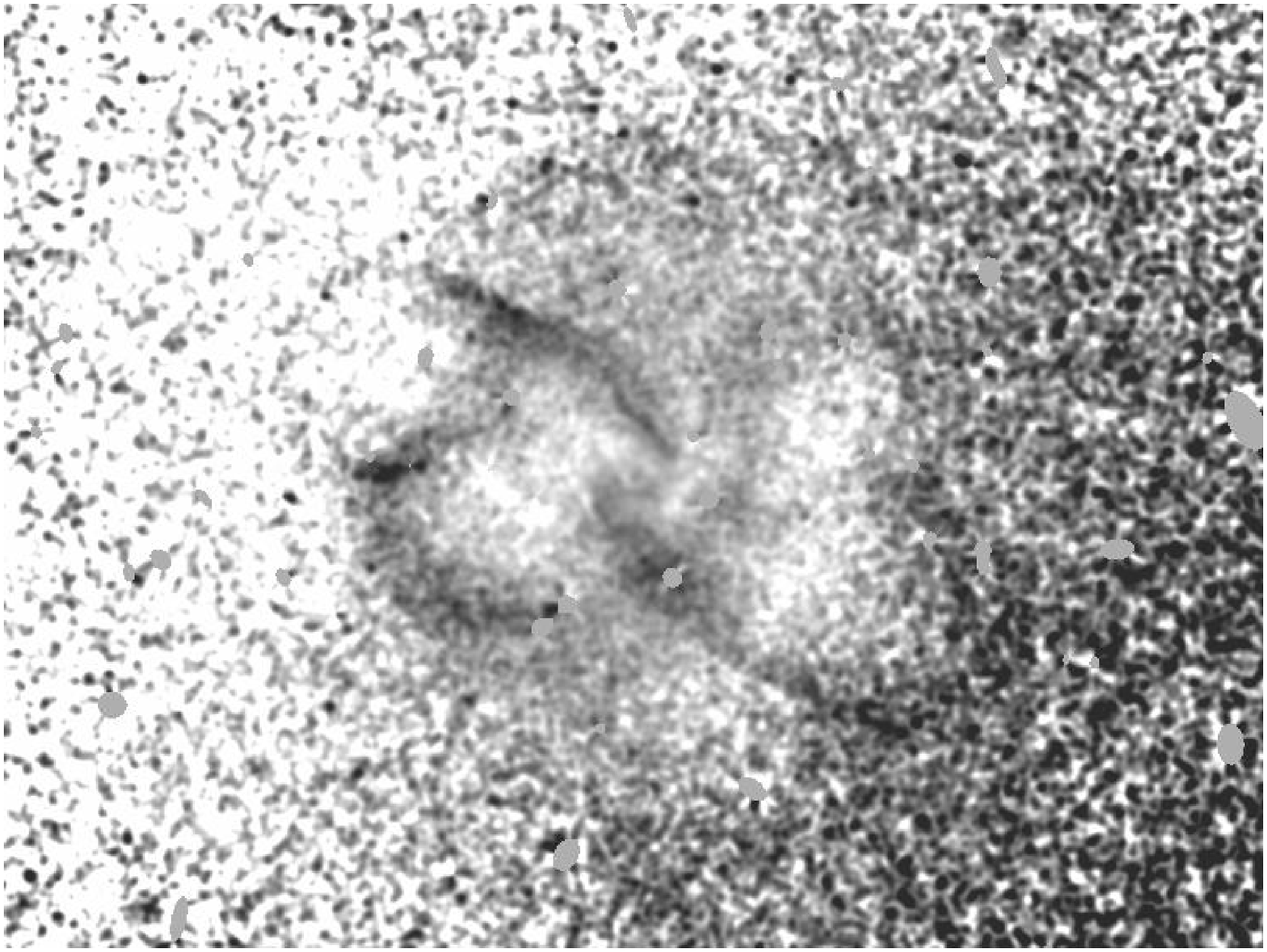}
\caption{{\it Left:} Chandra ACIS-I+ACIS-S image of NGC~4636 in the 0.3-2 keV band. The three bubble-like features
detected are labelled and identified by cyan ellipses. The yellow arrows are pointing toward 
the detected rims of the bubbles.
{\it Right:} Chandra ACIS-I+ACIS-S image after the subtraction of a $\beta$-model fitted to
the general diffuse X-ray emission. 
\label{xrayimage}}
\end{figure}

NGC~4636 was observed twice by ACIS-S and twice by ACIS-I.
In our analysis, we use the longer ACIS-S observation (performed on 2000, January 26; ObsID: 323) 
and both the ACIS-I observations, performed one immediately after the other 
on 2003, February 14 (ObsID: 3926) and on 2003, February 15 (ObsiD: 4415).
Details on data preparation and analysis can be found in \citet{baldi09}.
Fig.~\ref{xrayimage}a shows the merged 0.3-2~keV Chandra image of all three observations.
The galaxy presents a very bright central core of radius $\sim1~kpc$ (15$^{\prime\prime}$) centered on
the nucleus.
A lower surface brightness region surrounds the nucleus extending out to $\sim6$~kpc ($80^{\prime\prime}$).
Two quasi-symmetric arm-like features are embedded in this lower surface brightness emission. These features were 
previously reported by \citet{jones02} who analyzed the shorter ACIS-S observation. However, combining
these data with the two ACIS-I observations we are able to observe that the south-western arm is clearly
part of an X-ray cavity extending as far as $\sim9$~kpc ($2^{\prime}$) from the center. 
The hint of a similar structure can 
be observed in coincidence with the north-eastern arm-like feature. It is not clear from the X-ray
image whether there is a cavity in this case.
The cavities are more evident if we remove the contribution from the general diffuse
emission of the galaxy. Fig.~\ref{xrayimage}b shows the merged Chandra image in the 0.3-2 keV band where
a $\beta$-model fitted to the galaxy diffuse emission was subtracted. This processing allowed to highlight
fainter structures.

\section{The X-ray Bubbles}
\begin{figure}
\includegraphics[width=6cm]{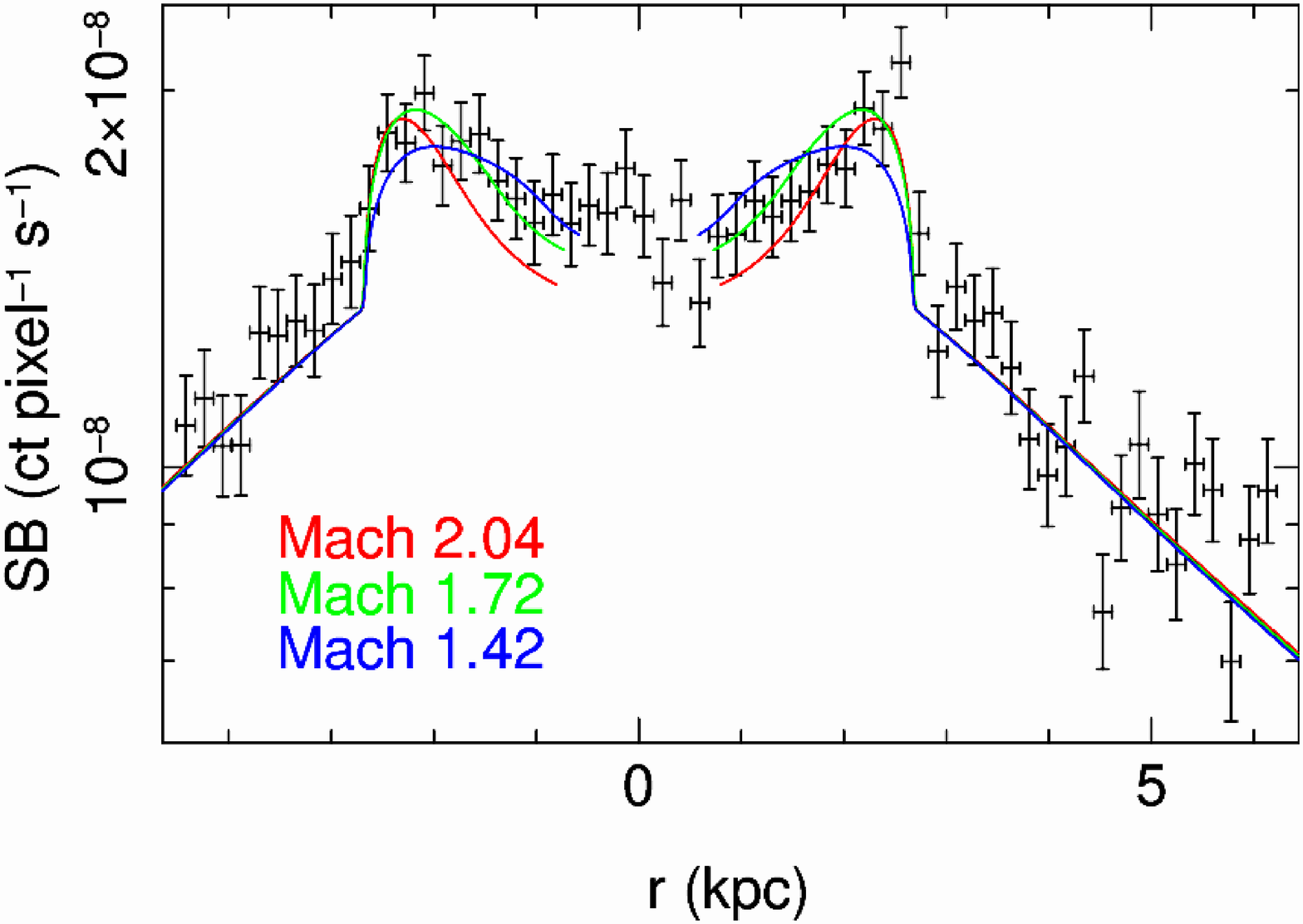}
\includegraphics[width=6cm]{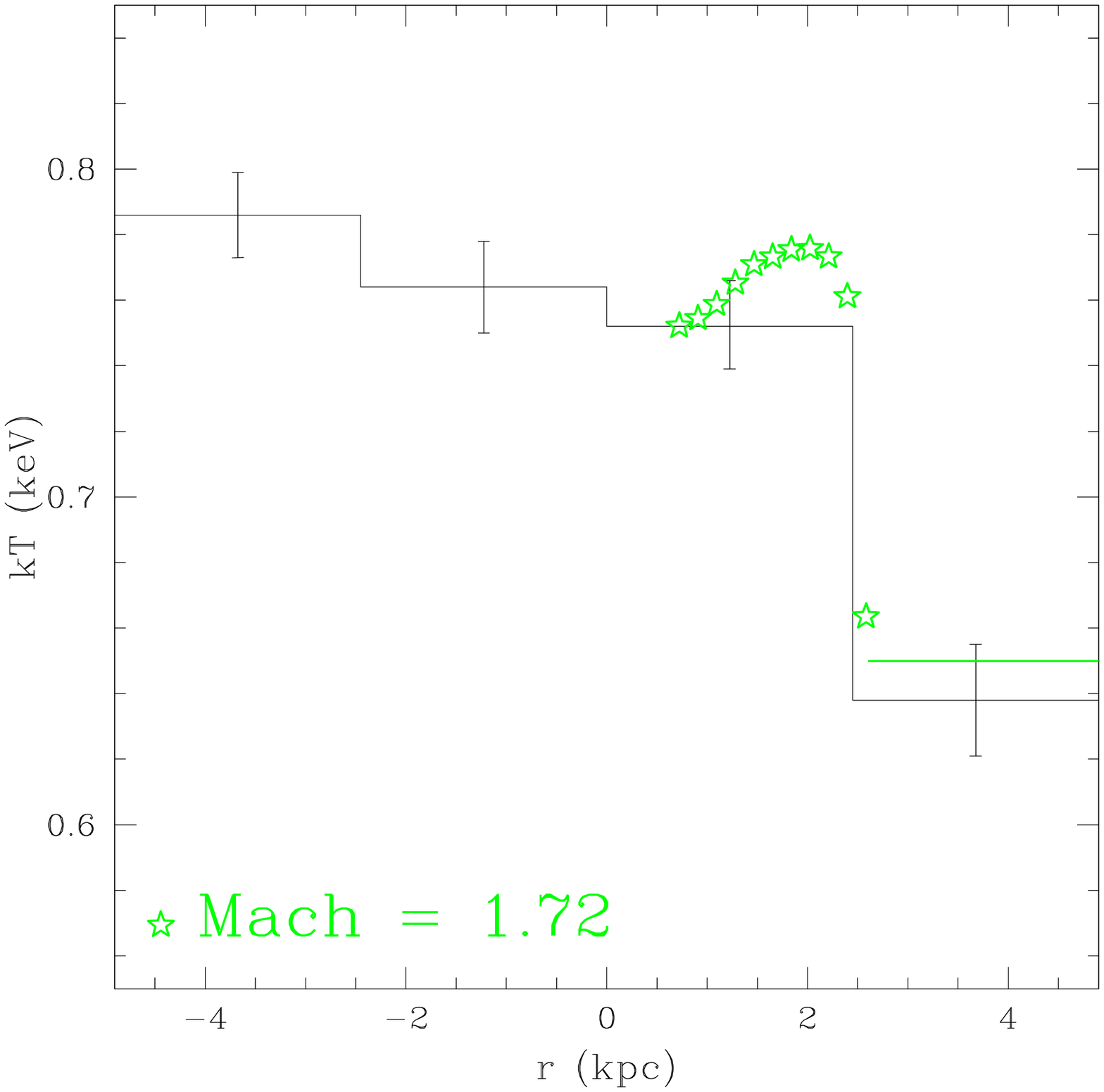}
\caption{{\it Left:} Surface brightness profile perpendicular to the shock front for the SW bubble. The three
colored solid lines represents the prediction from a numerical hydro-dynamical shock model at different
Mach numbers for the shock. The best fit shock model to the observed data has $M=1.72$. {\it Right:} 
Temperature profile across the southern rim of the SW bubble. The temperature jump is consistent with
the predictions of a shock model with $M=1.72$.\label{SWshock}}
\end{figure}

The most prominent features observed in the Chandra image are the two quasi symmetric 8~kpc
long X-ray arm-like structures, already observed by \citet{jones02} and by \citet{osullivan05}. The SW arm (SW1) 
is however part of a cavity
which extends at least $\sim5$~kpc in radius to the North.
We performed a spectral analysis in a strip perpendicular to the SW arm-like feature, dividing
the strip into rectangles $\sim0.5^\prime$ wide, 
to look for variations in temperature or metal abundance. 
Although the metallicity does not vary significantly
across the bubble rim, a sharp variation in the temperature was detected coincident with SW1 
(Fig.~\ref{SWshock}) 
showing a temperature decrease from $kT\sim0.75$~keV to $kT\sim0.64$~keV, well above
the measurement errors ($\sim0.01$~keV). A temperature jump was not observed across SW2 most likely because
of the complicated geometry of the X-ray emission. Indeed, SW2 seems to be partially embedded in another
bubble-like feature located just North of it.\\
The symmetry of the X-ray arm-like features is highly suggestive of the presence of a symmetric bubble NE of the nucleus
of NGC~4636. However the southern rim of the bubble is not clearly visible and it looks instead to be embedded
in another round shaped bubble located to the East of the nucleus. If we examine the surface brightness profile of the
NE cavity we find a shape which is very similar to the SW bubble. Performing a spectral analysis across the
northern cavity rim NE1, 
we also observe a 20\% temperature jump across the rim. However,
the scenario in this part of the galaxy is more complex because of the presence of an additional feature just East of the
nucleus with the shape of another cavity. This cavity looks less elongated than the other two cavities observed.
However, the surface brightness profile has a shape similar to the one across the NE cavity, while the temperature profile shows
a temperature jump similar to the one observed in the other two cavities.

\subsection{The origin of the cavities: a simple shock model}
The most likely scenario for the origin of the cavities observed in the X-ray morphology of NGC~4636
is that they were the result of successive outbursts of the central AGN.
In this scenario the jet propagated rapidly from the center, creating a long thin cavity which then inflated in
all directions. Perpendicular to the axis of the jet, the expansion has an approximate cylindrical symmetry.
The expanding cavities drive shocks into the surrounding gas.

A 1-dimensional, cylindrically symmetric, time-dependent hydrodynamic
model was used to investigate the properties of the shocks for the
SW bubble.
In this model the sound speed in the relativistic gas that fills the cavity (the piston that drives the shock) 
is assumed very high, keeping the pressure in the piston nearly uniform. The ratio of the pre-shock
pressure to the post-shock pressure determines the strength of the shock. As a result, the shock is weakest 
(slowest) in the region close to the AGN and fastest in the region farthest from the AGN. 
Further details on the shock model can be found in \citet{baldi09}.

The calculations of the physical parameters derived from the model were performed for the SW bubble.
The three cavities present similar temperature jumps and surface brightness profiles, so similar
physical parameters are expected.
From the hydrodynamic model, the age of the shock, $t\sim2\times10^{6}$~yrs, is the time it takes
to expand to its observed size. The age is only sensitive
to the shock strength and the temperature of the unshocked gas. This
age is slightly shorter than the ratio of the shock radius to the
present shock speed, because in the model the shock strength decreases
with time (i.e. the shock expanded faster when it was younger).
The total energy which produced the shock was $10^{56}$~ergs, roughly equal to the enthalpy, 
$H=4pV=7\times10^{55}$~ergs, calculated in the assumption that the bubble is predominantly relativistic 
($\gamma=4/3$). The average mechanical power required to produce the bubble equals to 
$P_{mech}\sim1.6\times10^{42}$~erg s$^{-1}$.

\bibliographystyle{aipproc}   

\bibliography{alesbib.bib}

\IfFileExists{\jobname.bbl}{}
 {\typeout{}
  \typeout{******************************************}
  \typeout{** Please run "bibtex \jobname" to optain}
  \typeout{** the bibliography and then re-run LaTeX}
  \typeout{** twice to fix the references!}
  \typeout{******************************************}
  \typeout{}
 }

\end{document}